\documentclass[twocolumn]{revtex4}
\usepackage{graphicx}
\usepackage{dcolumn}
\usepackage{bm}
\usepackage{color}
\usepackage[squaren]{SIunits}
\usepackage{amssymb}
\usepackage{natbib}

\begin{document}

\newcommand{\ie}{{\it i.e.}}
\newcommand{\eg}{{\it e.g.}}
\newcommand{\etal}{{\it et al.}}


\title{Superconductivity and Magnetic Properties of high-quality single crystals of $A_{x}$Fe$_2$Se$_2$ ($A$ = K and Cs)}

\author{J. J. Ying, X. F. Wang, X. G. Luo$^\dag$, A. F. Wang, M. Zhang, Y. J. Yan, Z. J. Xiang, R. H. Liu, P. Cheng, G. J. Ye and  X. H. Chen}
\altaffiliation{E-mail of X.H.C: chenxh@ustc.edu.cn\\$^\dag$ E-mail of X.G.L: xgluo@mail.ustc.edu.cn}
\affiliation{Hefei National Laboratory for Physical Science at
Microscale and Department of Physics, University of Science and
Technology of China, Hefei, Anhui 230026, People's Republic of
China}

\date{\today}


\begin{abstract}

We successfully grew the high-quality single crystals of
$A_{x}$Fe$_2$Se$_2$ ($A$ = K and Cs) by self-flux method. Sharp
superconducting transition was observed for both types of crystals.
The crystals show the onset superconducting transition temperatures
($T_{\rm c}$) of  31 K and 30 K for K- and Cs-compounds,
respectively, with nearly 100$\%$ shielding fraction. The crystals
show quite high resistivity in the normal state of more than 160
m$\Omega$ cm and 1300 m$\Omega$ cm maximum resistivity for
$K_{0.86}Fe_2Se_{1.82}$ and $Cs_{0.86}Fe_{1.66}Se_{2}$ single
crystals, respectively. Much larger upper critical field $H_{\rm
c2}$ is inferred from low-temperature iso-magnetic-field
magnetoresistance in these crystals than in FeSe. The anisotropy
$H^{ab}_{\rm c2}$(0)/$H^{c}_{\rm c2}$(0) is around 3 for both of the
two materials. Anisotropic peculiar magnetic behavior in normal
state has been found for $Cs_{0.86}Fe_{1.66}Se_{2}$

\end{abstract}

\maketitle

The layered iron-pnictide compounds have attracted the intense interests since superconductivity
at 26 K inZrCuSiAs-type LaFeAs(O,F)~\cite{Kamihara} was found. Replacement La with Sm leads to
superconducting transition temperature $T_{\rm c}$ = 43 K~\cite{chenxh} and soon renewed to highest
 record ($T_{\rm c}$ = 55 K)~\cite{ZARen,liurh} in this type of compound.  Up to now, various Fe-based
 superconductors, such as ZrCuSiAs-type $Ln$FeAsO ($Ln$ is rare earth elements)~\cite{Kamihara,chenxh,ZARen},
ThCr$_2$Si$_2$-type $Ae$Fe$_2$As$_2$ ($Ae$ is alkali earth
elements)~\cite{rotter1}, Fe$_2$As-type $A$FeAs ($A$ is Li or
Na)~\cite{CQJin, CWChu, Clarke} and anti-PbO-type
Fe(Se,Te)\cite{MKWu}, have been reported. The high $T_{\rm c}$ and
superconductivity proximity to a magnetically ordering
state\cite{PCDai,hchen}, which is thought to be similar to
high-$T_{\rm c}$ superconductor cuprates, inspired worldwide passion
of study towards elucidating the mechanism of high-$T_{\rm c}$
superconductivity. All of them have a common structural feature,
that is, the edge-sharing FeAs$_4$ (FeSe$_4$) tetrahedra formed FeAs
(FeSe) layers. The superconductivity in these compounds is thought
to be intimately related to the height of anion from Fe layer
~\cite{Mizuguchi1}. Unlike the case of FeAs-based compounds, which
usually possess cations between the FeAs layers, Fe(Se,Te) family
has a extremely simple structure with only FeSe layers stacked along
{\sl c}-axis without any intercalating cations.\cite{MKWu} To tune
the height of anion from Fe layer in Fe(Se,Te) can only be realized
by changing the relative proportion of Se and Te anions or by
applying high pressure. $T_{c}$ can reach 37 K (onset) under 4.5 GPa
from the ambient 8 K in FeSe .\cite{cava} The corresponding pressure
dependent ratio of $T_{\rm c}$ can reach as large as d$T_{\rm
c}$/d$P$ of $\sim$ 9.1 K/GPa, which is the highest among all the
Fe-base superconductors.\cite{cava} Takeing into account the effect
of local structure on superconductivity in the FeAs-based compounds,
superconductivity in Fe-Se family is expected to have a higher $T_c$
at ambient pressure by intercalating cations into between the FeSe
layers. Very recently, by intercalating K and Cs into between the
FeSe layers, superconductivity has been enhanced to be 30 K and 27 K
(onset temperature) without any external pressure in Fe-Se
system.\cite{xlchen,Mizuguchi,Krzton} Although the highest $T_{\rm
c}$ at ambient pressure for Fe-chalcogenides was achieved in
K-intercalated FeSe, the superconducting fraction is low and the
transition is broad. No full shielding fraction can be observed in
Cs-intercalated FeSe samples either. Therefore, to investigate
further intrinsic properties of intercalated FeSe compounds, single
crystals with high quality with full shielding fraction and sharp
transition are required to grow in priority.

In this report, we successfully grew the single crystals of
K$_x$Fe$_2$Se$_2$ and Cs$_x$Fe$_2$Se$_2$ with full shielding
fraction by self-flux method. The crystals showed the onset $T_{\rm
c}$ of 31 K and 30 K for K- and Cs-compounds, respectively. Nearly
100$\%$ superconducting volume fraction was observed through the
zero-field-cooling (ZFC) magnetic susceptibility measurements. Since
Fe-base high-$T_{\rm c}$ superconductors are thought to be related
to magnetic interaction closely, the normal-state magnetization was
also investigated.

\begin{figure}
\includegraphics[width = 0.45\textwidth]{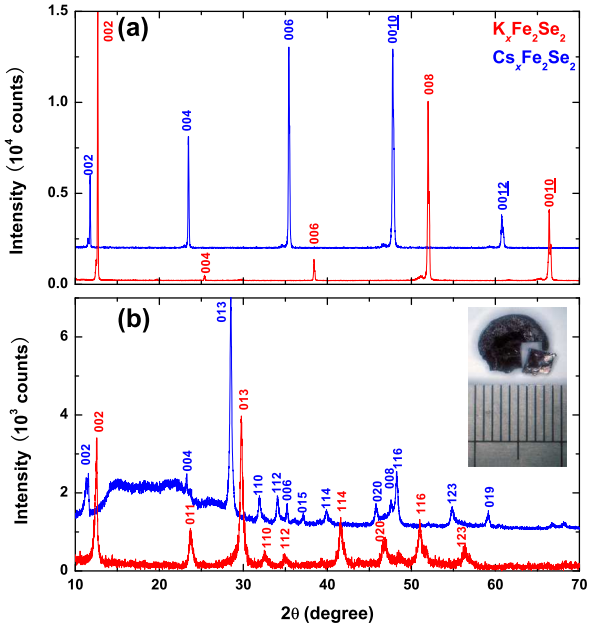}
\caption{(Color online) (a): The single crystal x-ray diffraction
pattern of K$_x$Fe$_2$Se$_2$ (red line) and Cs$_x$Fe$_2$Se$_2$ (blue
line); (b): X-ray diffraction pattern of the powdered
K$_x$Fe$_2$Se$_2$ (red) and Cs$_x$Fe$_2$Se$_2$ (blue line), all the
peaks can be well indexed. The hump in XDR pattern for
Cs$_x$Fe$_2$Se$_2$ comes from the Mylar film, which has been used to
protect the sample from air.}
\end{figure}

Single crystals AFe$_2$Se$_2$ (A=K, Cs) were grown by self-flux
method. Starting material FeSe was obtained by reacting Fe powder
with Se powder with Fe: Se = 1: 1 at 700$\celsius$ for 4 hours. K
and Cs pieces and FeSe powder were put into a small quartz tube with
nominal composition as K$_{0.8}$Fe$_2$Se$_2$ and
Cs$_{0.8}$Fe$_2$Se$_2$. Due to the high activity of K and Cs metal,
the single wall quartz tube will be corrupted and broken during the
growth procedure. Therefore, two wall quartz tube is required. We
realized it in the following way: the small quartz tube was sealed
under high vacuum and then was put in a bigger quartz tube following
by evacuating and being sealed. The mixture was heated to 1030
$\celsius$ in 4 hours and then kept at this temperature for 2 hours,
and later slowly cooled down to 750 $\celsius$ with 6 $\celsius$/
hours. After that, the temperature was cooled down to room
temperature by shutting down the furnace. The obtained single
crystals show the flat shiny surface with dark black color. The
inset of Fig.1b shows a piece of typical single crystal of
Cs$_x$Fe$_2$Se$_2$. For K-compound crystals are easy to cleave and
thin crystals with thickness less than 100 $\mu$m can be easily
obtained, while for Cs-compound crystals are fragile and quite
difficult to cleave.

Single crystals of K$_x$Fe$_2$Se$_2$ and Cs$_x$Fe$_2$Se$_2$ were
characterized by powder X-ray diffraction (XRD), and X-ray single
crystal diffraction, and Energy dispersive X-ray (EDX) spectroscopy,
and direct current (dc) magnetic susceptibility, and electrical
transport measurements. Powder XRD and single crystal XRD were
performed on TTRAX3 theta/theta rotating anode X-ray Diffractometer
(Japan) with Cu K$\alpha$ radiation and a fixed graphite
monochromator. Magnetic susceptibility is measured with the {\sl
Quantum Design} MPMS-SQUID. The measurement of resistivity and
magnetoresistance were performed using the {\sl Quantum Design}
PPMS-9.

\begin{figure}[ht]
\includegraphics[width = 0.45\textwidth]{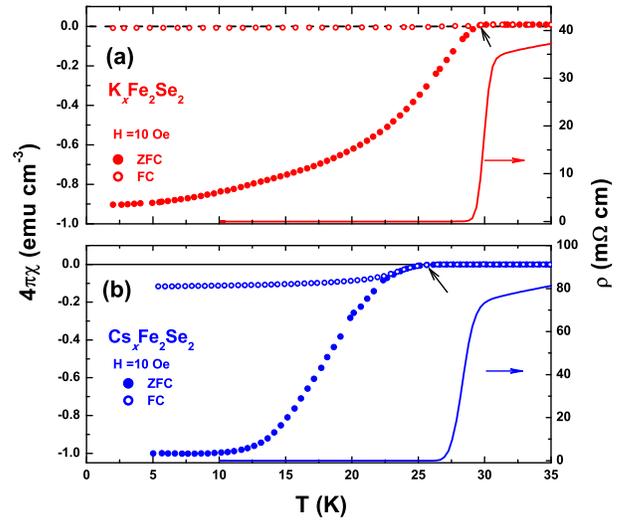}
\caption{(Color online) (a): The resistivity of
$K_{0.86}Fe_2Se_{1.82}$ around $T_{\rm c}$ (red solid line), and its
ZFC and FC susceptibility taken at 10 Oe with the magnetic field
parallel to the {\sl ab}-plane.  (b): The resistivity of
$Cs_{0.86}Fe_{1.66}Se_{2}$ around $T_{\rm c}$ (blue solid line), and
its ZFC and FC susceptibility taken at 10 Oe with the magnetic field
parallel to the {\sl ab}-plane.}
\end{figure}

Figure 1 shows the X-ray single crystal diffraction  and powder XRD
 after crushing the single crystals to powder for K$_x$Fe$_2$Se$_2$
and Cs$_x$Fe$_2$Se$_2$. Only (00$l$) reflections were recognized in
Fig. 1a, indicating that both types of single crystals were
perfectly oriented along {\sl c}-axis.  The lattice parameters are
obtained to index the powder XRD patterns in Fig. 1b with the
symmetry of I4/mmm and  $a$ = 3.8912 ${\rm \AA}$ and $c$ = 14.1390
${\rm \AA}$ for K$_x$Fe$_2$Se$_2$. The lattice constant $a$ obtained
here is smaller than that in Refs.14, 15 and 16. The actual
compositions of $K_{0.86}Fe_2Se_{1.82}$ were determined by EDX using
an average of different 6 points, indicating the existence of K and
Se deficiencies. This is different from previous reports, where K
and Fe sites show deficiencies.\cite{xlchen,Mizuguchi,Krzton}  For
Cs$_x$Fe$_2$Se$_2$, the lattice parameters determined from the
powder XRD patterns are $a$ = 3.9618 ${\rm \AA}$ and $c$ = 15.285
${\rm \AA}$, which is almost the same as that in Ref. 16. The actual
composition of $Cs_{0.86}Fe_{1.66}Se_{2}$ were determined by an
average of 6 different points EDX measurements, indicative of
deficiencies on both Cs and Fe sites. This is consistent with
previous reports.\cite{Mizuguchi,Krzton}

Figure 2a shows the resistivity and susceptibility measurements at
low temperature for $K_{0.86}Fe_2Se_{1.82}$. The resistivity starts
to drop quickly at about 31 K and it reaches zero at about 28.7 K.
The ZFC and field cooling (FC) susceptibilities show that the
superconducting shield begins to emerge at about 29 K with magnetic
field parallel to the {\sl ab}-plane taken at 10 Oe. The
superconducting volume fraction estimated from the ZFC magnetization
at 2 K is about 90$\%$, which indicates the bulk superconductivity
nature and good quality of the crystals. This superconducting
fraction is much larger than that in previous
reports.\cite{xlchen,Krzton,Mizuguchi} Fig.2b shows the
low-temperature resistivity and susceptibility measurements for
$Cs_{0.86}Fe_{1.66}Se_{2}$. The onset transition temperature is
about 30 K, and zero resistance is reached at 26.5 K. The
superconductivity transition temperature taken from the ZFC and FC
curves is estimated about 25.5 K. The superconducting volume
fraction estimated from the ZFC magnetization reaches 100$\%$ below
10 K estimated from the ZFC curve. It indicates a much better
superconductivity than the crystals in the previous
report.\cite{Krzton}

\begin{figure}[ht]
\includegraphics[width = 0.45\textwidth]{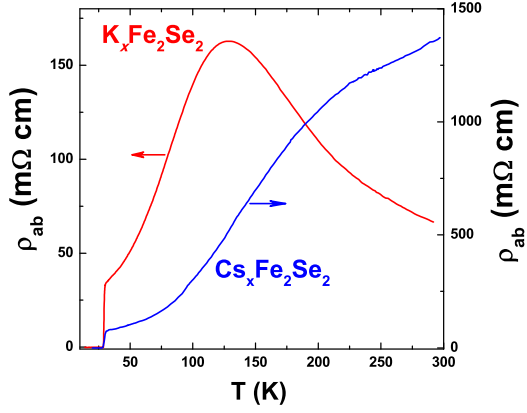}
\caption{(Color online) The temperature dependence of resistivity
for $K_{0.86}Fe_2Se_{1.82}$ (red line) and
$Cs_{0.86}Fe_{1.66}Se_{2}$ (blue line).}
\end{figure}

The temperature dependence of in-plane resistivity for the
$K_{0.86}Fe_2Se_{1.82}$ and $Cs_{0.86}Fe_{1.66}Se_{2}$ from 10 to
300 K. The $K_{0.86}Fe_2Se_{1.82}$ crystal shows the semiconducting
behavior at the high temperature and displays a broad maximum at
about 125 K. This temperature is slight higher than that in the
report by Guo {\it et al.} (around 100 K)~\cite{xlchen}, but much
smaller than that reported by Mizuguchi {\it et al.} ($\sim$200
K).\cite{Mizuguchi} Also, the maximum resistivity in our sample (170
m$\Omega$ cm) is much smaller than that in previous report
($\sim$3000 m$\Omega$ cm).\cite{Mizuguchi} The temperature and
magnitude of the maximum resistivity could be related to the
deficiency of Fe or Se, and large Fe deficiencies have been reported
by early authors.\cite{Mizuguchi,xlchen} With further decreasing the
temperature, the resistivity exhibits a metallic behavior and
superconductivity emerges at about 30 K. The resistivity of
$Cs_{0.86}Fe_{1.66}Se_{2}$ shows a metallic behavior in the whole
measuring temperature range, being quite different from that of the
$K_{0.86}Fe_2Se_{1.82}$. The resistivity of $K_{0.86}Fe_2Se_{1.82}$
and $Cs_{0.86}Fe_{1.66}Se_{2}$ are 70 m$\Omega$ cm and 1350
m$\Omega$ cm at room temperature, respectively, and all these values
are much larger than those of FeSe single crystals
\cite{Braithwaite} and the iron-pnictide superconductors\cite{wang}.
This may arise from the strong scattering from large disorder
induced by the Fe or Se deficiencies.

\begin{figure}[ht]
\includegraphics[width = 0.48\textwidth]{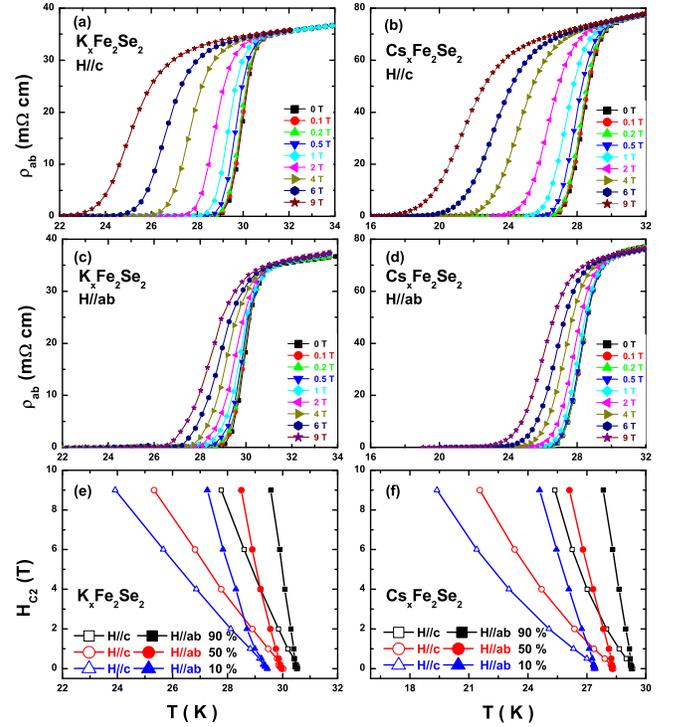}
\caption{(Color online) (a) and (c) show the temperature dependence
of resistivity for $K_{0.86}Fe_2Se_{1.82}$ crystal with the magnetic
field  parallel and perpendicular to the {\sl c}-axis respectively;
(b) and (d) show the temperature dependence of resistivity for
$Cs_{0.86}Fe_{1.66}Se_{2}$  crystal with the magnetic field parallel
and perpendicular to the {\sl c}-axis respectively; (e) and (f) show
the temperature dependence of $H_{\rm c2}$(T) for
$K_{0.86}Fe_2Se_{1.82}$ and $Cs_{0.86}Fe_{1.66}Se_{2}$,
respectively.}
\end{figure}

\begin{figure}[ht]
\includegraphics[width = 0.45\textwidth]{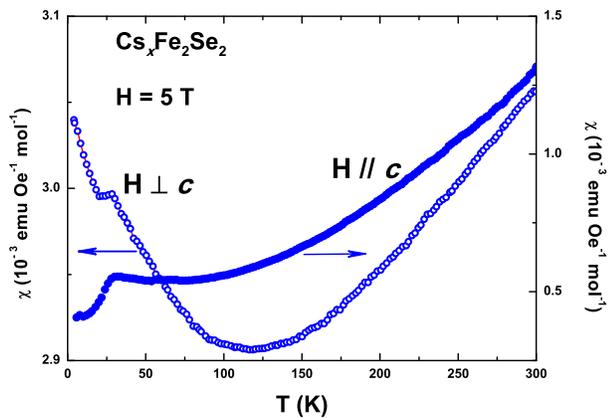}
\caption{(Color online) The magnetic susceptibility at 5 T for
$Cs_{0.86}Fe_{1.66}Se_{2}$ crsyatl with the magnetic field along and
perpendicular to {\sl c}-axis.}
\end{figure}

The resistivity of $K_{0.86}Fe_2Se_{1.82}$ and
$Cs_{0.86}Fe_{1.66}Se_{2}$ with the different magnitude of magnet
field applied parallel and perpendicular to the {\sl ab}-plane
around $T_{\rm c}$ are shown in Fig.5a, b, c and d, respectively.
The transition temperature of superconductivity is suppressed
gradually and the transition is broadened with increasing the
magnetic field . We defined the $T_{\rm c}$ as the temperature where
the resistivity was 90\%, 50\% and 10\% drop above the
superconducting transition. The anisotropy of the $H_{\rm c2}$($T$)
for $K_{0.86}Fe_2Se_{1.82}$ and $Cs_{0.86}Fe_{1.66}Se_{2}$ is shown
in Fig5.(e) and (f), respectively. Within the weak-coupling BCS
theory, the upper critical field at $T$=0 K can be determined by the
Werthamer-Helfand-Hohenberg (WHH) equation\cite{Werthamer} $H_{\rm
c2}(0)=0.693[-(dH_{\rm c2}/dT)]_{T_{\rm c}}T_{\rm c}$. Using the
data of $H_{\rm c2}$(T) for the 90$\%$ resistivity drop, one can
take $[-(dH_{\rm c2}^{ab}/dT)]_{T_{\rm c}}$ = 9.86 T/K, $[-(dH_{\rm
c2}^{c}/dT)]_{T_{\rm c}}$ = 3.17 T/K and $T_{\rm c}$ = 30.5 K. The
$H_{\rm c2}(0)$ can be estimated to be 208 T and 67 T with the field
parallel and perpendicular to the {\sl ab}-plane respectively for
$K_{0.86}Fe_2Se_{1.82}$, respectively. These values are a little
larger than those in previous report.\cite{Mizuguchi} For
$Cs_{0.86}Fe_{1.66}Se_{2}$ crystal, we take $[-(dH_{\rm
c2}^{ab}/dT)]_{T_{\rm c}}$ = 6.21 T/K, $[-(dH_{\rm
c2}^{c}/dT)]_{T_{\rm c}}$ = 2.13 T/K and $T_{\rm c}$ = 29.3 K. The
$H_{c2}(0)$ is 126 T and 43 T with the field parallel and
perpendicular to the {\sl ab}-plane, respectively. They are much
smaller than those of $K_{0.86}Fe_2Se_{1.82}$ although their $T_{\rm
c}$s are close to each other. The anisotropy $H^{ab}_{\rm
c2}$(0)/$H^{c}_{\rm c2}$(0) is about 3.1 and 2.9 for
$K_{0.86}Fe_2Se_{1.82}$ and $Cs_{0.86}Fe_{1.66}Se_{2}$,
respectively. This anisotropy value is larger than 1.70$\sim$1.86 in
$Ba_{0.60}K_{0.40}Fe_2As_2$\cite{wenhh}, but smaller than 4$\sim$6
in F-doped NdFeAsO\cite{Jia}.

Fig.5 shows the magnetic susceptibility of
$Cs_{0.86}Fe_{1.66}Se_{2}$ with the magnetic field of 5 T applied
parallel and perpendicular to the {\sl c}-axis. The magnetic
susceptibility decreases gradually with decreasing the temperature
at high temperature. The susceptibility starts to go upward at
around 120 K with the field perpendicular to the {\sl c}-axis. While
for the field parallel to the {\sl c}-axis, the susceptibility only
shows very tiny upturn at around 70 K. The anomaly around 26 K is
due to the transition of superconductivity. The susceptibility is
larger as field is applied in {\sl ab}-plane, which may suggest the
spins lies within the plane. The peculiar magnetic behavior
Cs$_x$F$e_2$Se$_2$ may be related to the deficiencies of Fe or the
occurrence of the superconductivity. The decrease in susceptibility
with decreasing temperature at high temperatures indicates there
exists an antiferromagnetic coupling. The occurrence of
superconductivity could be related to such peculiar magnetic
properties. The detailed magnetic
 structure requires further experiments.

In summary, we successfully grew the single crystals of K- and
Cs-intercalated FeSe compounds. $T_{\rm c}^{\rm onset}$ is 31 and 30
K for $K_{0.86}Fe_2Se_{1.82}$ and $Cs_{0.86}Fe_{1.66}Se_{2}$,
respectively. The ZFC dc magnetic susceptibility indicates that the
superconducting fraction is close to 100$\%$ for both types of
crystal. The large $H_{\rm c2}$ observed in these materials is
similar to the other iron-pnictide superconductors\cite{Yuan}. The
anisotropy $H^{ab}_{\rm c2}$(0)/$H^{c}_{\rm c2}$(0) is around 3 for
both the two materials. The normal state resistivity is very large
compared with the other iron-based superconductors. Such large
resistivity could arise from the Fe or Se vacancy in conducting FeSe
layer. Very high electrical resistance in these superconducting
compounds could challenge the theoretical scenario for the mechanism
of superconductivity. Anisotropic peculiar magnetic behavior has
been found in $Cs_{0.86}Fe_{1.66}Se_{2}$ crystal. This maybe related
to the deficiencies of Fe. It should be addressed that their $T_c$
is nearly the same although the resistivity behavior in
$K_{0.86}Fe_2Se_{1.82}$ and $Cs_{0.86}Fe_{1.66}Se_{2}$ is quite
different.
\\

{\bf ACKNOWLEDGEMENT} This work is supported by the Natural Science Foundation of China and by the Ministry of Science
and Technology of China, and by Chinese Academy of Sciences.\\

\end{document}